\documentstyle[11pt,newpasp,twoside,epsf]{article}
\def\edcomment#1{\iffalse\marginpar{\raggedright\sl#1\/}\else\relax\fi}
\reversemarginpar

\begin{document}
\title{The Circumstellar Matter Around OH44.8--2.3}
\author{M. Bobrowsky}
\affil{Challenger Center for Space Science Education, 1250 North Pitt Street, Alexandria, VA  22314}
\author{B.W. Greeley}
\affil{Orbital Sciences Corporation, 7500 Greenway Center Drive, Greenbelt, MD  20770}
\author{T. Ueta, M.M. Meixner}
\affil{Department of Astronomy, University of Illinois, Urbana, IL  61801}

\begin{abstract}
I-band images of the OH maser source OH44.8--2.3 are presented.  
Having a well-determined distance, this object is shown to be 
very young and physically small, with an asymmetric compact 
bipolar nebula already present.  In the morphological 
terminology of Meixner et al.\ (2000) and Ueta et al.\ (2000), 
this object would be classified as a DUPLEX nebula because of 
its high optical depth and pole-to-equator density contrast.

\end{abstract}

\section{Introduction}

OH/IR stars are important for understanding mass loss and stellar
evolution at the end of the AGB phase.  Mass loss rates have been
derived in a variety of ways:  from the IRAS 60 $\mu$m flux (Jura
1987), the ratio of the 25 $\mu$m to 12 $\mu$m flux (Van der Veen 1989),
the OH maser flux (Baud \& Habing 1983), and from an OH
photodissociation model combined with the spatial extent of the
OH maser emission.  Excluding the last of these, there is
reasonable agreement about the mass loss rate of OH44.8--2.3:  
M $\sim$10$^{-5}$ M$_\odot$ yr$^{-1}$.  Indeed, Heske et al.\ (1990) 
found that in the less
extreme OH/IR stars, like OH44.8--2.3, the mass loss rate derived
from infrared properties agrees reasonably well with that estimated
from the CO emission.

OH44.8 is highly variable (IRAS variability index 9) and has a
visual extinction of 1.9 magnitudes.  The total flux at Earth,
corrected for extinction, is 8.9 x 10$^{-11}$ W m$^{-2}$, and 
the luminosity is
3950 L$_\odot$ (Groenewegen 1994).  Groenewegen used a dust radiative
transfer model to simultaneously solve the radiative transfer
equation and the equation of thermal equilibrium for the dust
assuming a grain radius of 0.05 $\mu$m, a grain density of 2 g cm$^{-3}$, and
a condensation temperature of 1000 K.  Parameters derived from
the model included the optical depth at 5 $\mu$m ($\tau$ = 1.22), the inner
radius of the dust shell (r$_{inner}$ = 6.28 R$_*$), and the gas-to-dust ratio
($\psi$ = 0.0070).

\section{Observations}

HST observations were obtained with the WFPC2 Planetary
Camera in June and July 1999.  Dithered images were acquired
using the V (F555W) and I (F814W) filters.  The  dithered images
were interlaced according to the statistical significance of each
pixel using the ``drizzling" algorithm.  During the drizzling process,
the dithered frames were subpixelized resulting in a final image
with a higher resolution and smaller pixel 
size (0.$^{\prime\prime}$0228/pixel).
Bad pixels (due to cosmic-rays) were removed by replacing the
pixels with a median of the neighboring pixels in each of the
dithered frames.  Deconvolution was carried out using the Lucy
algorithm.

\section{Results}

OH44.8--2.3 is not visible in the V-band images, but it can be
clearly seen in the I-band images shown both before (Fig. 1) and
after (Fig. 2) deconvolution.  
\begin{figure}
\plotfiddle{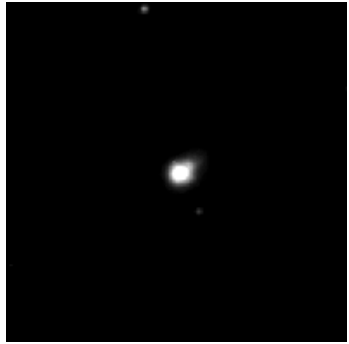}{3.0in}{0.0}{65.}{65.}{-200}{-150}
\caption{HST WFPC2 PC image of OH44.8--2.3 acquired with the I (F814W) filter.}
\end{figure}

\begin{figure}
\plotfiddle{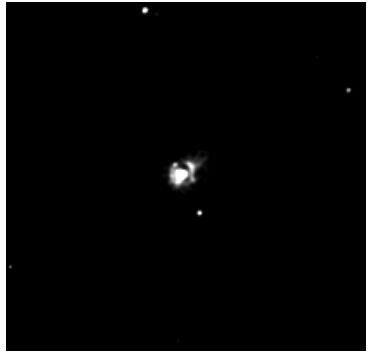}{3.0in}{0.0}{65.}{65.}{-200}{-150}
\caption{I-band (F814W) image of OH44.8--2.3 with Lucy deconvolution.}
\end{figure}
It is extremely compact, which is not
surprising since it was unresolved in the mid-infrared survey of
Meixner et al.\ (1999).  The spatial extent of OH44.8 
is $\sim$0.$^{\prime\prime}$4 x
0.$^{\prime\prime}$6, with the long axis oriented approximately 
north-south.  The
deconvolved image shows a toroidal structure surrounding
apparent bipolar outflows.  At a distance of 1.2 kpc (van
Langevelde et al.\ 1990), the radius of the torus is 2.3 x 10$^{15}$ cm.
An expansion velocity of 16 km s$^{-1}$ (te Lintel Hekkert 1990, Meyer
et al.\ 1998) leads to an age of only 45 yr!

OH44.8 would be classified as a DUPLEX object in the
classification scheme of Meixner et al.\ (2000) and Ueta et al.\
(2000).  With a totally obscured central star, this object is similar to
IRAS 16342--3814, an extreme AGB star, also known to have H$_2$O
and OH maser emission (Likkel \& Morris 1988, Sahai et al.\ 1999).
Also like IRAS 16342--3814, OH44.8 has unequal bipolar lobes in
which some inner structure can be seen.  DUPLEX nebulae like
that seen in OH44.8 might be expected to be younger than SOLE
nebulae.  This would be consistent with both the young age of
OH44.8 and its late spectral type.  (The late spectral type also
probably accounts for why OH44.8 was not seen in the V-band
images.)  Finally, it appears that bipolar nebulae are preferentially
found close to the Galactic plane (Corradi \& Schwarz (1995),
consistent with OH44.8's Galactic height of 100 pc.  Thus, besides
being an OH maser source, OH44.8 is a very young and very dusty
bipolar nebula.

\section{Acknowledgments}

This research is based on observations with the NASA/ESA Hubble 
Space Telescope, obtained at the Space Telescope Science Institute, 
which is operated by the Association of Universities for Research 
in Astronomy, Inc. Under NASA contract No.\ NAS5-26555.  Bobrowsky is 
supported by NASA Grant GO-06364.01-94A, and Meixner is supported 
by NASA Grant GO-06364.02-95A.

\end{document}